\documentclass[prb,10pt,superscriptaddress,twocolumn]{revtex4}
\usepackage{ccfonts}

\usepackage{amsfonts}
\usepackage{amssymb}
\usepackage{psfrag} \usepackage{graphicx}

\begin{document}

\title{Anomalous Dimension of Dirac's Gauge-Invariant Nonlocal Order
  Parameter in Ginzburg-Landau Field Theory}

\author{H. Kleinert}
\email{kleinert@physik.fu-berlin.de}
\affiliation{Institut f\"ur Theoretische Physik, Freie Universit\"at
  Berlin, Arnimallee 14, D-14195 Berlin, Germany }

\author{Adriaan M. J. Schakel}
\email{schakel@itp.uni-leipzig.de}
\affiliation{Institut f\"ur Theoretische Physik, Universit\"at Leipzig,
  Augustusplatz 10/11, D-04109 Leipzig, Germany}

\date{\today}

\begin{abstract}
  In a Ginzburg-Landau theory with $n$ fields, the anomalous dimension
  of the gauge-invariant nonlocal order parameter defined by the
  long-distance limit of Dirac's gauge-invariant two-point function is
  calculated.  The result is exact for all $n$ to first order in
  $\epsilon \equiv 4-d$, and for all $d\in (2,4)$ to first order in
  $1/n$, and coincides with the previously calculated gauge-dependent
  exponent in the Landau gauge.
\end{abstract}

\maketitle

\section{Introduction}

An outstanding problem in gauge theories is the construction of physical
correlation functions or propagators of the charged matter fields.  As
such objects involve fields located at different points in spacetime,
the standard forms, expressed solely in terms of the matter fields, are
in general not gauge invariant and, consequently, not physical.  The
principle of gauge invariance by itself does not yield a unique
prescription, and various solutions have been proposed a long
time ago, notably by Dirac \cite{Dirac} and by Schwinger
\cite{Schwinger}.  These proposals have been used to investigate
important physical problems such as anomalies, quark potentials, and
order parameters distinguishing the different phases of gauge theories.
Recently, the issue has received considerable attention in the context
of high-temperature superconductors, where massless Dirac fermions
coupled to a dynamical gauge field were put forward as an effective
theory for studying the unusual properties of the normal state of
underdoped materials \cite{RantnerWen}.

In this Letter we contribute to this issue by showing that different
gauge-invariant proposals for correlation functions lead to different
physical results.  We do so by considering the abelian Higgs or
Ginzburg-Landau model, which describes a great variety of physical
systems, ranging from scalar QED, superconductors over liquid crystals
to cosmic strings and vortex lines in superfluids \cite{GFCM}.  The
model consists of a $|\phi|^4$-theory coupled minimally to the
electromagnetic gauge field $A_\mu$, with $\mu= 1, \cdots d$. Its
Hamiltonian is
\begin{equation}
  \label{H}
  \mathcal{H} \!=\! \left|(\partial_\mu \!+ \!i e A_\mu)\phi\right|^2 + m^2
  |\phi|^2 + \lambda |\phi|^4 + \frac{1}{4} F^2_{\mu \nu}+ \frac{1}{2
    \alpha} (\partial_\mu A_\mu)^2 ,
\end{equation}
where $F_{\mu \nu} = \partial_\mu A_\nu - \partial_\nu A_\mu$.  The
scalar field $\phi$ has $n/2$ complex components and a O($n$)-symmetric
self-interaction with coupling constant $\lambda$.  The coefficients $e$
and $m$ denote electric charge and mass parameters of the complex $\phi$
field, respectively.  The last term with parameter $\alpha$ fixes a
Lorentz-invariant gauge.  We use mostly the notation of statistical
field theory in $d$ space dimensions.  The results apply, however,
equally to quantum field theory in $d$ spacetime dimensions in the
Euclidean formulation.

In the following, we will work at criticality by setting $m=0$.  The
free correlation function $G(x-x') = \langle \phi(x) \phi^\dagger(x')
\rangle_0$ of the scalar field is the Fourier transform of $1/k^2$:
\begin{equation}
  G(x) = \int \frac{\mathrm{d}^d k}{(2 \pi)^d} \frac{\mathrm{e}^{\mathrm{i} k \cdot
      x}}{k^2} = \frac{\Gamma(d/2-1)}{4 \pi^{d/2}} \frac{1}{x^{d-2}}.
  \label{@FCF}
\end{equation}
For the free correlation function $D_{\mu \nu }(x-x') = \langle A_\mu(x)
A_\nu (x') \rangle_0$ of the gauge field, we must Fourier transform
\begin{equation}
  \label{aprop}
  D_{\mu \nu}(q) = \frac{1}{q^2} \left[ \delta_{\mu \nu} - (1-\alpha)
    \frac{q_\mu q_\nu}{q^2} \right] ,
\end{equation}
and obtain
\begin{eqnarray} \!\!\!\!\!\!\!\!\!\!
  \label{aprop_real}
  D_{\mu \nu}(x) = \frac{\Gamma(d/2\!-\!1)}{8 \pi^{d/2}x^{d-2}}
  \left[(1\!+\!\alpha) \delta_{\mu \nu} + (d\!-\!2)(1\!-\! \alpha) \frac{x_\mu
      x_\nu}{x^2} \right].  \!\!\!\!\!\!\!\!\!\!\!\!\!\!\!\!   \nonumber \\
\end{eqnarray}

\begin{figure}
\begin{center}
\includegraphics[width=3.0cm,height=5.cm,angle=-90]{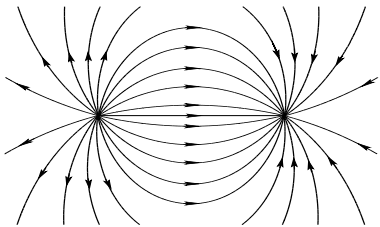}
\end{center}
\caption{Flow lines corresponding to the Dirac proposal for the external current $J_\mu(z)$
  specified by Eqs.~(\ref{@EQ1}) in two dimensions.
  \label{fig:dipole}}
\end{figure}

In the presence of interactions, the expectation value $\langle \phi(x)
\phi^\dagger(x') \rangle$ is an unphysical quantity since it is not
invariant under gauge transformations
\begin{equation}
  \phi(x) \to \mathrm{e}^{\mathrm{i} e \Lambda (x)} \phi(x), \quad A_\mu (x) \to
  A_\mu (x) - \partial_\mu \Lambda (x).
\end{equation}
In fact, it vanishes identically due to Elitzur's theorem \cite{EL}.  A
gauge-invariant correlation function was first proposed by Dirac
\cite{Dirac}.  Adapted to our purposes, it reads
\begin{equation}
  \label{indep}
  \mathsf{G}(x-x') = \left\langle \phi(x) \,
    \mathrm{e}^{\mathrm{i} e\int \mathrm{d}^d z J_\mu(z)A_\mu(z)}
    \phi^\dagger(x')
  \right\rangle.
\end{equation}
The average denoted by angle brackets is taken with respect to the full
Hamiltonian (\ref{H}), and the external current $J_\mu(z)$ satisfies the
equations
\begin{equation}
  \partial_\mu J_\mu(z) = \delta(z-x') -  \delta(z-x), \quad
  \partial ^2J_\mu(z) =0,
  \label{@EQ1}
\end{equation}
where the first ensures the conservation of the external current in the
presence of a source of strength $+1$ at $x$ and a sink of strength $-1$
at $x'$ \cite{footnote1}.  The explicit form of the external current
(see Fig.~\ref{fig:dipole}) is $J_\mu(z) = J'_\mu(z-x') - J'_\mu(z-x)$,
where
\begin{equation}
  J'_\mu(z) = -i \int \frac{\mathrm{d}^d k}{(2 \pi)^d} \frac{k_\mu}{k^2}
  \mathrm{e}^{i k \cdot z} =  -  \frac{\Gamma(d/2-1)}{4 \pi^{d/2}}
  \partial_\mu \frac{1}{z^{d-2}}.
  \label{@SO}
\end{equation}
Being nonlocal, the current in Eq.~(\ref{indep}) is more properly
denoted by $J_\mu(z;x,x')$.  At the critical point, the gauge-invariant
correlation function (\ref{indep}) is expected to have the power
behavior
\begin{equation}
  \label{algebra2}
  \mathsf{G}(x) \sim \frac{1}{x^{d-2+\eta_{\rm GI}}},
\end{equation}
with the Fisher exponent $\eta_{\rm GI}$.  In the ordered phase, the
correlation function (\ref{indep}) has the large-distance behavior
\begin{equation}
  \mathsf{G}(x-x') \mathop{\longrightarrow }_{|x-x'|\rightarrow
  \infty}|\tilde \phi| ^2,
  \label{@ORD}
\end{equation}
where $\tilde \phi(x)$ is the nonlocal order parameter
\begin{equation}
  \tilde \phi(x)\equiv \mathrm{e}^{-\mathrm{i}e\int \mathrm{d}^d z \,
    J'_\mu(z-x)A_\mu(z)}\phi(x).
  \label{@}
\end{equation}
Since $J'_\mu(z)$ is a total derivative [see Eq.~(\ref{@SO})], $\tilde
\phi(x)$ reduces after a partial integration in the Landau gauge
$\partial_\mu A_\mu=0$ to the local form $\phi(x)$ \cite{King}.  In
other words, $J_\mu$ becomes invisible in this gauge and the value for
$\eta_{\rm GI}$ is expected to coincide with the gauge-dependent result
for $\langle \phi(x) \phi^\dagger(x') \rangle$ obtained in the gauge
$\alpha=0$.

The purpose of this Letter is to determine $\eta_{\rm GI}$ to first
order in $\epsilon\equiv 4-d$ and also for all $d \in (2,4)$ to first
order in $1/n$.

Note that of the two equations in (\ref{@EQ1}), only the source
equation is needed for gauge invariance of (\ref{indep}).  This can also
be solved by the $\delta$-function on a line $L$ running from $x'$ to
$x$,
\begin{equation}
  J_\mu(z)= \delta _\mu(z;L)\equiv \int_L \mathrm{d} s \,
  \frac{\mathrm{d}{\bar x}_\mu(s)}{\mathrm{d} s} \delta (z-\bar x(s)).
  \label{@}
\end{equation}
For a straight line $L$ with $\bar x_\mu(s)=x'(1-s)+sx,\,s\in [0,1]$,
this leads to Schwinger's gauge-invariant correlation function
\cite{Schwinger}
\begin{equation}
  \label{indep1}
  \left\langle \phi(x) \,
    \mathrm{e}^{-\mathrm{i} e \int_{x}^{x'} \mathrm{d} \bar{x}_\mu A_\mu(\bar{x})}
    \phi^\dagger(x')
  \right\rangle,
\end{equation}
whose critical properties we studied in Ref.~[\onlinecite{KLSCH}].  The
Schwinger construction can be thought of having all the external current
originating from the source at $x$ and terminating at the sink at $x'$
squeezed into an infinitely thin line along the shortest path connecting
the two points.  In the disordered phase, the current lines have a
finite line tension, and this correlation function vanishes
exponentially for large distances.  Because the finite line tension
exponentially suppresses larger loops, only a few small current loops
are present in this phase.  Upon approaching the critical point, the
line tension vanishes and current loops can grow without energy cost.
Their proliferation signals the onset of superconductivity
\cite{KS,CAM,KKS,leshouches}.

The dependence of Eq.~(\ref{indep1}) on the shape of $L$ can also be
seen more formally.  We observe that a deformation of $L$ is a new type
of gauge transformation discussed extensively in Refs.~[\onlinecite{GFCM,CAM}]:
\begin{equation}
 \delta _\mu(z;L)\rightarrow
 \delta _\mu(z;L')=
 \delta _\mu(z;L)+\partial _ \nu  \delta _{\mu \nu }(z;S),
\label{@}
\end{equation}
where $\delta _{ \mu\nu }(z;S)$ is the $\delta $-function on the surface
$S$ swept out in the deformation $L\rightarrow L'$.  Under this gauge
transformation, the correlation function (\ref{indep1}) changes by a
nontrivial phase
\begin{equation}
  \bar \phi(x)
  \mathop{\longrightarrow }_{L\rightarrow L'}
  \mathrm{e}^{-\mathrm{i}e\int d^dz  F_{\mu \nu }(z)
    \delta _{\mu \nu }(z;S)}\bar\phi(x).
\label{@FIE3}\end{equation}

Due to the different physical content of the correlation functions
involved, the critical behavior to be derived here for Dirac's
correlation function (\ref{indep}) will be quite different from that of
Schwinger's (\ref{indep1}) calculated in our previous note
[\onlinecite{KLSCH}].

\section{$ \epsilon $-expansion}
Perturbation theory yields via Wick's theorem, three perturbative
corrections to (\ref{indep}) to lowest order in $e^2$:
\begin{equation}
  \label{TTT}
  \mathsf{G}(x-x') = G + T_0 + T_1 + T_2,
\end{equation}
where $T_{0,1,2}$ contain zero, one, and two factors of the flow field
$J_\mu$.

The term $T_0$ is calculated by standard methods.  Infrared divergences
are avoided by evaluating Feynman diagrams at a finite external momentum
$\kappa$.  Being the only scale available, $\kappa$ is used to render
dimensionful parameters such as $e^2$ dimensionless: $\hat{e}^2 = e^2
\kappa^{d-4}$.  The result is the well-known gauge-dependent
contribution \cite{HLM,Kang}:
\begin{equation}
  \label{eta0}
  \eta_\phi
  = \frac{\alpha - 3}{8 \pi^2} \hat{e}^2_* = \frac{6}{n}
  (\alpha -3)\epsilon.
\end{equation}
The lowest-order $\epsilon$-expansion on the right hand is obtained by
inserting for $\hat e^2$ the charge $\hat{e}^2_* = 48 \pi^2 \epsilon/n$
at the infrared-stable fixed point, which at one loop exists only for
$n>12(15 + 4 \sqrt{15}) \approx 365.9$.
At two loop,
different resummation techniques suggest the
existence of a fixed point for
the physical case $n=2$ \cite{FoHo,KNO}.

Next, we calculate the last term in Eq.~(\ref{TTT}):
\begin{equation}
  \label{T2}
  T_2(x-x')\! =\! - \frac{e^2}{2} G(x-x')\! \int\! \mathrm{d}^d z \mathrm{d}^d z'
  J_\mu (z) D_{\mu \nu}(z-z') J_\nu (z')
\end{equation}
which splits in a separate scalar and gauge part.  Several integrations
by part reduce the integrals in coordinate spacetime to the generic form
\begin{eqnarray}
  \label{integral}
  &&\!\!\!\!\!\!\!\!\!\!\!\!\!\!\!\int \mathrm{d}^d z
  \frac{1}{|z-x|^{d-2}} \frac{1}{|z-x'|^{d-p}}
  \nonumber \\&&\!\!\!\!\!\! =
  \frac{2 \pi^{d/2}}{p}  \frac{\Gamma(d/2-1-p/2)}{\Gamma(d/2-1)
    \Gamma(d/2-p/2)} \frac{1}{|x-x'|^{d-2-p}},
\end{eqnarray}
and we obtain, with the abbreviations $\partial_\mu \equiv
\partial/\partial x_\mu, \partial'_\mu \equiv \partial/\partial x'_\mu$,
\begin{eqnarray}
  \lefteqn{T_2(x-x') = \frac{e^2}{64 \pi^{3d/2}} \Gamma^3(d/2-1)  G(x-x')}
  \\ && \!\!\!\!\!\!\!\! \times \left(
    \partial_\mu \partial'_\mu \int \mathrm{d}^d z
    \mathrm{d}^d z' \frac{1}{|z\!-\!x|^{d-2} |z\!-\!z'|^{d-2} |z'\!-\!x'|^{d-2}}
  \right. \nonumber \\&&\left.\!\!\!\!\!\!\!\!-\frac{1}{2} \frac{1\!-\!\alpha}{d\!-\!4} \partial^2
    {\partial'}^2 \!\int \mathrm{d}^d z \mathrm{d}^d z' \frac{1}{|z\!-\!x|^{d-2}
      |z\!-\!z'|^{d-4} |z'\!-\!x'|^{d-2}} \right)\!. \nonumber
\end{eqnarray}
In the limit of small $\epsilon =4-d$, this reduces to
\begin{equation}
  T_2(x-x') = - \hat{e}^2 \frac{\alpha}{8 \pi^2}  G(x-x') \ln(\kappa |x-x'|) .
\end{equation}
Comparison with an expansion of (\ref{algebra2}) in powers of $\eta_{\rm
  GI}$ gives a contribution to the Fisher exponent proportional to the
gauge-fixing parameter $\alpha$:
\begin{equation}
  \label{eta2}
  \eta_2 =  \frac{\alpha}{8 \pi^2} \hat{e}^2_* .
\end{equation}
This result can be checked by considering the ratio of the correlation
functions $\langle \phi(x) \phi^\dagger(x') \rangle$ and $\langle
\mathrm{e}^{-\mathrm{i} e \int \mathrm{d}^d z J_\mu(z)A_\mu(z)}
\rangle$.  Adapting an argument given in
Ref.~[\onlinecite{Khveshchenko_02}], one can show that this ratio and,
consequently, the combination $\eta_\phi - \eta_2$ should be independent
of the gauge-fixing parameter $\alpha$.  The expressions (\ref{eta0})
and (\ref{eta2}) indeed fulfill this requirement.  In our previous study
\cite{KLSCH}, we found for the Schwinger correlation function as only
difference an additional contribution to $\eta_2$ independent of
$\alpha$.

We are left with the calculation of the second, or mixed term in
Eq.~(\ref{TTT}), which reads explicitly
\begin{eqnarray}
  \label{T1}
  T_1(x-x') & =&e^2 \int \mathrm{d}^d z \, \mathrm{d}^d z' \left[G(x-z)
    \stackrel{\leftrightarrow}{\frac{\partial}{\partial z_\mu}} G(z-x') \right]
  \nonumber \\&&\times
  J_\nu(z') D_{\mu \nu}(z-z'),
\end{eqnarray}
where the right-minus-left derivative
$\stackrel{\leftrightarrow}{\partial}_\mu = \partial_\mu -
\stackrel{\leftarrow}{\partial}_\mu$ operates only within the square
brackets.  To logarithmic accuracy, we can write \cite{Khveshchenko_02}
\begin{eqnarray}
  T_1(x-x') &\approx& e\, G(x-x')
  \int \mathrm{d}^d z \, \mathrm{d}^d z'
  \left[\frac{\partial}{\partial z_\mu} G(z-x')
  \right.\nonumber \\&&\left. - \frac{\partial}{\partial
      z_\mu} G(x -z) \right] J_\nu(z') D_{\mu \nu}(z-z').
\end{eqnarray}
Both terms in the square brackets give the same contribution.
Proceeding in the same way as before, we find
\begin{equation}
  \label{T1T2}
  T_1 = -2 T_2,
\end{equation}
and therefore as contribution to the Fisher exponent
\begin{equation}
  \eta_1 = -2 \eta_2 = -  \frac{\alpha}{4 \pi^2} \hat{e}^2_* .
\end{equation}
This contribution, which is again proportional to the gauge-fixing
parameter $\alpha$, is identical to the one found for the Schwinger
correlation function \cite{KLSCH}.  Added together, we obtain for the
manifestly gauge-invariant correlation function (\ref{indep})
\begin{equation}
  \label{fi}
  \eta_\mathrm{GI} = \eta_\phi + \eta_1 + \eta_2 = - \frac{3}{8
    \pi^2} \hat{e}^2_* = - \frac{18}{n} \epsilon .
\end{equation}
As expected, this result for the nonlocal Dirac order parameter
coincides with the value for $\eta_\phi$ obtained in the Landau gauge
$(\alpha=0)$.  When $\alpha$ is considered a running coupling constant
of the theory (\ref{H}), the Landau gauge emerges as a fixed point
$\alpha_*=0$ of the renormalization group \cite{LAN}.  This is a special
case of the more general result \cite{LitimPawlowski} that $\alpha=0$
is always a fixed point when considering a gauge-fixing term of the form
$(L_\mu A_\mu)^2/2 \alpha $.  The choice $L_\mu = \partial_\mu$ then
leads to the Landau gauge, while the choice $L_\mu=n_\mu$, with $n_\mu$
a constant vector, leads to the axial gauge.

The expression (\ref{fi}) is to be contrasted with $\eta_\mathrm{GI} = -
(3/4 \pi^2) \hat{e}^2_*$ we \cite{KLSCH} derived for Schwinger's
gauge-invariant correlation function (\ref{indep1}).  It follows that
the exponent (\ref{fi}) characterizing Dirac's correlation function is
less negative than the one characterizing Schwinger's.  The latter
coincides with $\eta_\phi$ in the gauge $\alpha =-3$.  For this value of
$\alpha$, the external current line connecting $x$ to $x'$ has no
effect.  A similar observation in the context of quantum chromodynamics
was made in Ref.~[\onlinecite{CraigieDorn}].

\section{Large-$n$ expansion}

The leading contribution in $1/n$ generated by fluctuations in the gauge
field is obtained by dressing its correlation function with arbitrary
many bubble insertions, and adding the infinite set of Feynman graphs
\cite{AppelHeinz}.  The resulting geometric series leads to the
following change in the prefactor of the correlation function
(\ref{aprop}):
\begin{equation}
  \label{mod}
  \frac{1}{q^2} \to \frac{1}{q^2 + n e^2 [c(d)/2(d-1)]  \, q^{d-2} },
\end{equation}
where the second term in the denominator dominates the first for small
$q$ if $d \in (2,4)$.  The constant $c(d)$ stands for the 1-loop
integral
\begin{equation}
  \label{cd}
  c(d) \!=\! \int \frac{\mathrm{d}^d k}{(2 \pi)^d} \frac{1}{k^2 (k +p)^2}
  \biggr|_{p^2=1}\!\!\!\! = \frac{\Gamma(2-d/2) \Gamma^2(d/2-1)}{(4 \pi)^{d/2}
    \Gamma(d-2)},
\end{equation}
where analytic regularization is used as before to control
ultraviolet divergences.  To leading order in $1/n$, the value of
$\eta_\phi$ for $d\in (2,4)$ reads \cite{Hikami,VN}
\begin{equation}
  \label{largeN}
  \eta_\phi = \frac{2}{n} \frac{4-d -(d-1) [4(d-1) -d \, \alpha]}{(4
    \pi)^{d/2} c(d) \Gamma(d/2 +1)},
\end{equation}
which depends on the gauge-fixing parameter $\alpha$.  For
$d=4-\epsilon$, this result reduces to Eq.~(\ref{eta0}) obtained to
first order in $ \epsilon $.

We next consider the gauge-invariant version of this.  The term $T_2$ in
(\ref{T2}) can be evaluated as before.  To extract the dependence on
$\ln(|x-x'|)$ it will be useful to replace $q^{d-2}$ by $q^{d-2 +
  \delta}$ in Eq.~(\ref{mod}), with a dummy parameter $ \delta $, which
will be taken to zero at the end.  Then the large-$n$ limit of the
gauge-field correlation function becomes
\begin{equation}
  \label{apropN}
  D_{\mu \nu}(q) = \frac{2}{n e^2} \frac{d-1}{c(d)} \frac{1}{q^{d-2 +
      \delta}} \left[ \delta_{\mu \nu} - (1-\alpha) \frac{q_\mu q_\nu}{q^2}
  \right] ,
\end{equation}
or in coordinate spacetime
\begin{eqnarray}
  D_{\mu \nu}(x) &=& \frac{8}{n e^2}
  \frac{d-1}{c(d)}  \frac{1}{(4 \pi)^{d/2}
    \Gamma(d/2)}  \frac{1}{x^{2 - \delta}}
  \nonumber \\&\times&\left[\frac{1}{2} (d-3+\alpha) \delta_{\mu \nu} + (1- \alpha) \frac{x_\mu
      x_\nu}{x^2} \right].
\end{eqnarray}
Proceeding in the same way as before, we find after various
integrations by parts
\begin{eqnarray}
  \lefteqn{T_2(x-x') = \frac{1}{n} \frac{1}{2^{d+1} \pi^{3d/2}}
    \frac{d-1}{c(d)} \Gamma(d/2-1) G(x-x')} \\ &&\!\!\!\!\!\!\! \times \left( \partial_\mu
    \partial'_\mu \int \mathrm{d}^d z \mathrm{d}^d z' \frac{1}{|z\!-\!x|^{d-2}
      |z\!-\!z'|^{2-\delta} |z'\!-\!x'|^{d-2}}
  \right.\nonumber \\&&\!\!\!\!\!\!\! \left.\!+\!\hspace{1pt}  \frac{1}{d\!-\!2}
    \frac{1\!-\!\alpha}{\delta}
    \partial^2 {\partial'}^2 \!\!\!\int \!\!\mathrm{d}^d z \mathrm{d}^d z'
    \frac{1}{|z\!-\!x|^{d\!-\!2} |z\!-\!z'|^{\!-\!\delta} |z'\!-\!x'|^{d\!-\!2}} \right)\!. \nonumber
\end{eqnarray}
Using the integral formula (\ref{integral}), we obtain for $\eta_2$:
\begin{equation}
  \label{GI}
  \eta_2 = \alpha \frac{4}{n} \frac{d-1}{(4 \pi)^{d/2} c(d) \Gamma(d/2)} .
\end{equation}
This large-$n$ result valid for all $d\in (2,4)$ is once more
proportional to the gauge-fixing parameter $\alpha$, just as for small
$\epsilon$ in Eq.~(\ref{eta2}).  The result can again be easily checked
by noting that the combination $\eta_\phi - \eta_2$ is independent of
the gauge-fixing parameter $\alpha$.  As for small $\epsilon$, the
result for the Schwinger correlation function differs only by an
$\alpha$-independent contribution to $\eta_2$.

For the mixed term $T_1$, we also find for large $n$ the relation
(\ref{T1T2}) between the two contributions $T_1$ and $T_2$, and thus
$\eta_1 = -2 \eta_2$.  This expression for $\eta_1$ is identical to the
one for the Schwinger correlation function.  Adding the three
contributions together, we arrive at
\begin{equation}
  \label{grand}
  \eta_\mathrm{GI} = \frac{2}{n} \frac{(7 - 4 d) d}{(4
    \pi)^{d/2} c(d) \Gamma(d/2 +1)},
\end{equation}
independent of the gauge-fixing parameter $\alpha$.  This result, valid
for all $d\in (2,4)$, is the leading contribution in $1/n$.  As
expected, it coincides with the value (\ref{largeN}) for $\eta_\phi$
obtained in the Landau gauge $(\alpha=0)$.  This should be compared to
the critical exponent for the Schwinger correlation function
\cite{KLSCH} (\ref{indep1}) which coincides with Eq.~(\ref{largeN})
obtained in the {\em traceless gauge} $\alpha =1-d$, of which
$\alpha=-3$ found in the $\epsilon$-expansion is a special case.  In
this gauge, where the correlation function $D_{\mu \nu}$ is traceless,
the external current line connecting $x$ to $x'$ becomes invisible.
Although less than for Schwinger's correlation function,
$\eta_\mathrm{GI}$ found here is negative for small $\epsilon$ and all
$n$ or for $d\in (2,4)$ and large $n$.  In a recent Monte Carlo study
\cite{Olsson} of the three-dimensional lattice model ($n=2$) in the
London limit where $|\phi|=\mbox{const}$, the large negative value
$\eta_\phi=-0.79(1)$ was obtained in the Landau gauge.

\begin{acknowledgments}

  We thank F. Nogueira for many discussions. H.K. also thanks the
  European Network COSLAB for partial support. A.S. is grateful to W.
  Janke for the kind hospitality at the ITP, University Leipzig.  The
  work by A.S. is partially supported by the DFG Grant No. JA 483/17-3
  and by the German-Israel Foundation (GIF) under Grant No.\
  I-653-181.14/1999.

\end{acknowledgments}


\begin{thebibliography}{99}
\bibitem{footnote1} When the model (\ref{H}) is viewed as a quantum
  field theory in $d$ spacetime dimensions, the source and sink in
  Eq.~(\ref{@EQ1}) correspond to an instanton and anti-instanton,
  respectively.  This is different from Dirac's original construction,
  where that equation only refers to the $d-1$ spatial components $J_i$
  of the current, taking the form $\partial_i J_i(z) = \delta(z-x') -
  \delta(z-x)$ and $\partial_i^2 J_j(z) = 0$.  The source and sink then
  represent electric charges which generate an electric dipole field,
  that is needed to satisfy Gauss's law.
\end{thebibliography}
\end{document}